\documentclass[a4paper,UKenglish,cleveref, autoref, thm-restate,authorcolumns,usenames,dvipsnames]{lipics-v2019}
\usepackage{booktabs} 
\usepackage{mathpartir}
\usepackage{epigraph}
\usepackage{xspace}
\usepackage{listings}
\usepackage{graphicx}
\usepackage{tikz}
\usepackage{amsmath}
\usepackage{amssymb}
\usepackage{stmaryrd}

\usepackage{arydshln}
\newcommand{\specline}[1]{{\color{blue}\left\{#1\right\}}}

\newcommand{\proofoutstepnr}[5][1pt/3pt]{ \left.
\begin{array}{@{}l@{}}
\hdashline[#1]
\ensuremath{\specline{#2}}\\
\begin{array}{@{\;\;\;}l}
#3
\end{array}\\
\ensuremath{\specline{#5}}\\
\hdashline[1pt/3pt]
\end{array}
\color{OliveGreen}\right)\hspace{-6pt}{\color{OliveGreen}-}
\begin{array}{@{}l@{}}
#4
\end{array}
}
\newcommand{\proofoutstep}[5][1pt/3pt]{
\proofoutstepnr[#1]{#2}
{#3}
{\rotatebox[origin=c]{90}{$\begin{array}{@{}c@{}}{}#4\end{array}$}}
{#5}
}

\nolinenumbers

\newcommand{\readable}{\texttt{\textcolor{blue}{readable}}\xspace}
\newcommand{\writable}{\texttt{\textcolor{blue}{writable}}\xspace}
\newcommand{\isolated}{\texttt{\textcolor{blue}{isolated}}\xspace}
\newcommand{\immutable}{\texttt{\textcolor{blue}{immutable}}\xspace}

\newcommand{\db}[1]{\llbracket#1\rrbracket}

\newcommand{\result}[1]{\mathsf{y = \mathsf{result}}}

\newcommand{\assert}[1]{\{#1\}}

\definecolor{framed}{gray}{0.50}


\EventEditors{Robert Hirschfeld and Tobias Pape}
\EventNoEds{2}
\EventLongTitle{34th European Conference on Object-Oriented Programming (ECOOP 2020)}
\EventShortTitle{ECOOP 2020}
\EventAcronym{ECOOP}
\EventYear{2020}
\EventDate{July 13--17, 2020}
\EventLocation{Berlin, Germany}
\EventLogo{}
\SeriesVolume{166}
\ArticleNo{10}
\category{Pearl}
\hideLIPIcs

\title{Designing with Static Capabilities and Effects: Use, Mention, and Invariants}

\author{Colin S. Gordon}{Department of Computer Science, Drexel University, USA}{csgordon@drexel.edu}{0000-0002-9012-4490}{}
\authorrunning{C. S. Gordon}
\Copyright{Colin S. Gordon}

\begin{document}
\maketitle

\begin{abstract}
  Capabilities (whether object or reference capabilities) are fundamentally tools to restrict effects.  Thus static capabilities (object or reference) and effect systems take different technical machinery to the same core problem of statically restricting or reasoning about effects in programs.
  Any time two approaches can in principle address the same sets of problems, it becomes important to understand the trade-offs between the approaches, how these trade-offs might interact with the problem at hand.

  Experts who have worked in these areas tend to find the trade-offs somewhat obvious, having considered them in context before.  However, this kind of design discussion is often written down only implicitly as comparison between two approaches for a specific program reasoning problem, rather than as a discussion of general trade-offs between general classes of techniques.
  As a result, it is not uncommon to set out to solve a problem with one technique, only to find the other better-suited.

  We discuss the trade-offs between static capabilities (specifically reference capabilities) and effect systems, articulating the challenges each approach tends to have in isolation, and how these are sometimes mitigated.  We also put our discussion in context, by appealing to examples of how these trade-offs 
  were considered in the course of developing prior systems in the area.
  Along the way, we highlight how seemingly-minor aspects of type systems --- weakening/framing and the mere existence of type contexts --- play a subtle role in the efficacy of these systems.
\end{abstract}

\section{Introduction}

Capabilities are a classic idea~\cite{levy84capability,lampson1974protection} with intuitive appeal: explicitly tie possession of certain entities to the ability to perform certain actions, so by bounding the flow of those entities one can restrict the possible actions of a program or program component~\cite{RobustComposition}.
Much of the work in this area centers the notion of \emph{object capabilities}, where capabilities control access to objects (in the OO sense), and capabilities are realized as object references: a program fragment cannot modify or invoke operations of an object it cannot reference.  This immediately grants a way to control mutation of objects, and by tying external calls to specific objects, also extends to controlling externally-visible behaviors as well.   For example, by associating all file operations with a particular \emph{object} --- not a globally accessible library call --- developers may tightly control which code can access those operations by restricting how widely the file operations object is distributed.
Intuitively, capabilities act as permission to do things, and the absence of capabilities acts as a lack of permission.
It is also possible to delegate partial access to an object's operations using
proxy objects~\cite{Chase:1992:LSO:141936.141969,van2013trustworthy,VanCutsem:2010:PDP:1869631.1869638} or through capabilities acting as handles to trusted mediators~\cite{levy84capability}.
However, doing this kind of reasoning statically is also appealing, because it incurs no runtime performance overhead when delegating or mediating access.

As a result, there is now a rich body of work on statically checked capabilities.
Once static reasoning is employed, the kinds of restrictions proxies and mediators permit in object capability systems may not require new dynamic objects exposing different sets of operations.  One of the most well-developed bodies of work on static capabilities uses 
\emph{reference capabilities}, which associate different permissions\footnote{Typically reference capabilities are distinguished statically, though a dynamic interpretation is possible.} to individual references in a program, in contrast to the object capability view that all references to an object are equal and restrictions stem from using different objects with fewer or modified operations available.
Thus, different references to the same object --- distinguished by a type system or static analysis, but not the runtime system --- may permit programs holding them different abilities to affect object state or invoke certain operations.  
Most reference capability systems are type systems where reference types come equipped with a type qualifier~\cite{Foster1999} corresponding to certain permissions (and in some cases, invariants or assumptions about aliases).
Capability-based reasoning is supported by checking the types flowing into a given program context.

Different variations of reference capabilities have been employed to solve a wide array of programming problems.
Systems with read-only reference capabilities~\cite{tschantz05,zibin10,dietl07,noble1998flexible,clarke1998ownership} restrict some references to read-only access to their referents --- even when aliases exist that can be used to mutate the referent --- which is useful for preventing a wide variety of accidental mutations, from expressing that a method treats its arguments as deeply read-only to controlling consequences of representation exposure~\cite{detlefs}.  This can be combined nicely with object immutability~\cite{zibin07}, as all references to an immutable object are read-only.  Transitive versions have been used to ensure data race freedom in Microsoft prototypes~\cite{oopsla12} and the Pony programming language~\cite{clebsch2015deny}: if two threads only shared (transitively) read-only references, no data races can exist between them.\footnote{This is simplifying away some of the substructural aspects of these type systems, which all make use of forms of uniqueness to \emph{also} support partitioning mutable data between threads, or in Encore's case~\cite{CastegrenW17} restrict conflicting accesses to atomic synchronization primitives.}
They have also been used to infer method purity~\cite{Huang2012,huang2012inference}: if a method accepts only (transitively) read-only inputs (including the receiver), it has no externally-visible side effects.\footnote{This sets aside extensionally-observable effects, such as allocating memory or triggering GC.}
In other contexts, program behavior can be constrained by building more fine-grained capabilities that grant not only all-or-none permission to mutate, but can grant permission for only certain kinds of mutation, and can therefore enforce nuanced invariants by restricting which capabilities can coexist for the same resource~\cite{pldi13,dissertation,toplas17,militao14,Militao16Composing,CastegrenW16,CastegrenW17}.

Each of these systems makes critical use of the original motivation for capabilities: by restricting what flows into certain parts of the program, one can provide guarantees about what that part may do --- without precisely examining the semantics of its internals.

For the problems mentioned above, there also exist effect systems to statically check the same high level concepts (e.g., for data race freedom~\cite{bocchino09}, or purity~\cite{rytz2013flow}).  In contrast to capability systems which reason externally in terms of what capabilities flow into code, effect systems are a class of type system extensions that analyze program behavior\footnote{By effect systems, we mean the sort of type system extension that reasons about bounds on program behavior as part of the type judgment, in the sense of the work on FX that originally coined the term ``effect system''~\cite{lucassen88,gifford86}.  This stands in contrast to \emph{denotational} approaches which attempt to assign meaning to effects, often by way of monads or some extension thereof, following Moggi~\cite{Moggi1989}. Filinksi~\cite{Filinski2010} offers an excellent discussion of the distinction.  Readers familiar with algebraic effects should note that much work on algebraic effects involves \emph{both} varieties: handlers define the semantics of user-defined effects, but a restrictive type system of the sort we discuss ensures all effect operations are invoked in the context of an enclosing handler.} by (to a first approximation) performing a bottom-up analysis of what interesting actions might occur, based on (typically) a join semilattice of effects: primitive or external actions of interest are typed as having particular specific effects representing their behavior of interest, and larger expressions' effects are computed by taking the least upper bound of subexpressions' effects.
This raises a fundamental question: when considering a static reasoning approach to a problem, how do we recognize which approach is likely to be better suited?
Comparisons between these systems in the literature tend to focus on the low-level expressive distinctions between systems for a particular problem domain (e.g., System A accepts this data race free program rejected by System B), or the relative complexity of the type rules (as a proxy for usability).
While the core trade-offs are there if one looks carefully, the broader issue of contrasting the trade-offs between these \emph{classes} of solutions has received little explicit attention.

In our personal experience, 
it is not uncommon to set out to build a capability-based system, only to find
effect systems more suitable to the task at hand --- and sometimes the reverse.
People experienced with both effect systems and capability systems --- whether as designers, or users --- likely find this unsurprising.  But to the best of our knowledge, there is essentially no discussion in the literature on how system designers chose one approach over the other; systems are presented as complete and finished designs, then evaluated against other finished designs --- the design process is lost.
Having a record of these trade-offs and design questions would be useful, both for shared understanding and especially for newcomers to static capabilities or effects.  Part of this requires identifying and developing terminology for aspects of these trade-offs.

In this expository paper we articulate some of the core trade-offs between these static reasoning approaches, and how these trade-offs are moderated in important ways by some of the most humble of reasoning principles in type systems: weakening and the use of type contexts.  We also explain how the trade-offs have affected the design of 
several reference capability systems and effect systems we have worked on.
We expect that little of what we say would be surprising to those who have worked on both static reference capability systems and type-and-effect systems; we view our contributions as primarily giving clear explicit exposition to these trade-offs that are generally left implicit in the literature, and putting those trade-offs in context by providing some extra information on the design evolution of a couple effect systems and capability systems.
Our hope is that newcomers to these areas and their intersection, or outsiders looking in, will find these distinctions helpful.

\section{Capabilities, Use, and Mention}
One of the original goals for capability-based design is to reason about the effect of some code by reasoning about the capabilities it is provided~\cite[Ch. 8]{RobustComposition} --- a long-standing practice based on the notion that capabilities essentially grant permission to cause effects, though until relatively recently~\cite{craig2018,liu2020stoic,liu2016} the exact relationship between static capabilities and static effects was left implicit.\footnote{This despite being noted as a relevant question (in other terms) much earlier~\cite{Foster1999}.}
However, the pure form of this approach --- that the set of capabilities provided is used to give an upper bound on an expression's effect --- has limitations we have not seen crisply articulated in a general way before.

Before we can precisely articulate the limitations of a kind of reasoning, let us first clarify exactly what kind of reasoning we mean.
We will ground discussion primarily in systems using read-only references to control side effects~\cite{zibin07,zibin10,tschantz05} to control side effects.
We will draw on a prototype dialect of C\#~\cite{oopsla12} that used these and with context bounding to both control interference between threads, and also to strengthen typing assumptions.
These ideas were later generalized in Pony~\cite{clebsch2015deny}, so much of our discussion applies fairly directly there as well, and less precisely to a range of earlier~\cite{zibin07,zibin10,tschantz05}, contemporaneous~\cite{huang2012inference,Huang2012}, and later~\cite{giannini2019flexible} systems.
This line of work, focusing on read-only references devoid any meaning besides mutability restrictions, historically used the term \emph{reference immutability} to describe the relevant techniques (as in, an object was immutable \emph{through a particular (read-only) reference}). This was partly to distinguish itself from other techniques with read-only references as part of systems that captured more design intent, like owner-as-modifier ownership types~\cite{noble1998flexible,clarke1998ownership,clarke2013ownership} or universe types~\cite{dietl2005universes,dietl07}. To date, the particular sort of capability-bounding we discuss below has only been explored for systems in the so-called reference immutability family.

In most of these systems, the specific capability appears as a type qualifier~\cite{Foster1999} modifying a basic object (class) type.
In the C\# dialect we discuss, there were four reference capabilities: \isolated (externally unique~\cite{haller10}), \readable (transitively read-only), \writable (mutable), and \immutable (transitively immutable, in the sense that the immediate referent and all objects reachable from it are permanently immutable).
\readable and \immutable may be used for reading fields, may not be used for mutation, and may not be used to \emph{obtain} references usable for mutation.  For example, reading a \writable-declared field through a \readable reference would produce only a \readable reference --- e.g., iterating over the elements of a \lstinline|readable List<writable Foo>| would work through a series of \lstinline|readable Foo| instances.
In the case of an immutable referent, stronger results appear because everything reachable via an \immutable reference is an immutable object: iterating over a \lstinline|immutable List<writable Foo>| would see \lstinline|immutable Foo| instances.

Setting aside some subtleties related to uniqueness, the simplest embodiment of using context bounds to reason about effects in these systems is the parallel composition type rule from
the C\# dialect, present in an analogous form in related systems~\cite{clebsch2015deny,giannini2019flexible}:
 
\begin{mathpar}
\inferrule[T-Par]{
  \mathsf{NoWritable}(\Gamma_1,\Gamma_2)\\
  \Gamma_1\vdash C_1 \dashv\Gamma_1'\\
  \Gamma_2\vdash C_2 \dashv\Gamma_2'
}{
  \Gamma_1,\Gamma_2\vdash C_1 || C_2 \dashv \Gamma_1',\Gamma_2'
}
\end{mathpar}

The rule above simply says that as long as two thread bodies ($C_1$ and $C_2$) require no \writable variables in their inputs, it is safe (data-race-free) to run them in parallel (and the output of the flow-sensitive typing judgment combines per-thread outputs in the obvious way). This works because in order for a data race to occur, one thread would need write access to an object the other thread could reach.  Prohibiting \writable references from entering either thread guarantees this cannot happen: any object reachable from both threads would be truly immutable, or both threads would have only \readable references to it.
Crucially, this reasoning is sound because
the C\# dialect --- like Pony~\cite{clebsch2015deny}, Encore~\cite{CastegrenW17}, and L42~\cite{giannini2019flexible} --- prohibited global mutable state, providing a form of capability safety~\cite{RobustComposition}: assurance that reasoning in terms of only capabilities directly entering an expression was sound, because there were no \emph{ambient} capabilities (those that can be obtained by any code at any time).
This rule also partitions some mutable state between threads, by splitting up \isolated references.

Another, slightly less traditional form of effect-bounding is the notion of \emph{recovery}, first proposed by Gordon et al.~\cite{oopsla12}, adapted by Clebsch et al.~\cite{clebsch2015deny} for use in Pony, and later extended for better flexibility by Giannini et al.~\cite{giannini2019flexible}. Again, we will demonstrate with the simplest rule, one of two given for the C\# dialect:

\begin{mathpar}
\inferrule[T-RecoverImm]{
  \mathsf{IsolatedOrImmutable}(\Gamma)\\
  \mathsf{IsolatedOrImmutable}(\Gamma')\\
  \Gamma\vdash C \dashv \Gamma',x:\mathsf{readable}~D
}{
  \Gamma\vdash C \dashv \Gamma',x:\mathsf{immutable}~D
}
\end{mathpar}
This rule says that if all inputs and all but one output $x$ of a command are \isolated or \immutable, and the other output $x$ is \readable, then it is safe to recover the \emph{stronger} \immutable capability for $x$ --- stronger because \lstinline|immutable D| is a subtype of \lstinline|readable D|, making this a kind of statically safe downcast.  Intuitively this handles either the case that $x$ already points to immutable data, or the case that it points to mutable data that is unreachable except via $x$ and can therefore be ``frozen'' to immutable. The restrictions on $\Gamma$ and $\Gamma'$ ensure $x$ doesn't alias mutable state, since the lack of ambient capabilities means $x$ must point to an input (all \immutable or \isolated --- and therefore freezable) or something allocated within $C$ (which cannot escape via $C$'s inputs).
As with the concurrency rule, the soundness of this relies fundamentally on the fact that weak permissions on the inputs imposes strong restrictions on the code's behavior (plus the prohibition on ambient authority / global mutable state). 

This rule makes it possible to write code that handles some number of immutable or externally-unique data with code that was not written with strict immutability (as opposed to \readable) in mind:

\begin{lstlisting}
    readable T RandomChoice(readable T a, readable T b) { ... }
    ...
    `\assert{x,y:\immutable~T}`
    z = RandomChoice(x,y);
    `\assert{x,y:\immutable~T,z:\readable~T}`
    `\assert{x,y,z:\immutable~T}`
\end{lstlisting}

The code above passes two immutable references to \lstinline|RandomChoice|, which assumes it simply returns a \readable reference.  But with the recovery rule above, the result (\lstinline|z|) can be recovered as immutable --- it must either be pointer-equal to \lstinline|x| or \lstinline|y|, or a new \lstinline|T| allocated inside the method, which is therefore not aliased elsewhere, and can be converted to immutable.

In the C\# prototype, and now Pony, this kind of reasoning has worked out well.  But why, and where would it break?

\subsection{The Gap Between Capability Bounds and Effects: Use-Mention Distinction}
These kinds of reasoning could be done using explicit effect systems~\cite{lucassen88,gifford86}.  But what does that gain us?  As is known~\cite{melicher2017b,craig2018}, an explicit effect system requires a system that cares about the details of the code being analyzed, which can require complex types and effects~\cite{bocchino09} (we see examples in Section \ref{sec:effects}).  So concretely, what can effect systems offer that capability-based reasoning struggles with?

The key point of departure between this capability-bound-based reasoning and a general effect system is what we will refer to as a kind of \emph{use-mention distinction}.  In philosophy and linguistics, logical fallacies and confusion are known to arise from conflating \emph{use} of a thing with mere \emph{mention} of a thing~\cite{moore1986significant,davidson1979quotation}.
Reasoning about an expression's effect using only the capabilities it has access to inherently performs the same kind of conflation: possessing authority means only that the code has the \emph{ability} to use it, not that it necessarily does.
This seems to be anecdotally understood among designers of static capability systems, but rarely discussed.  To the best of our knowledge, this paper is the first to explicitly call out and name this trade-off.

Consider the following in the C\# reference immutability dialect:
\begin{lstlisting}
    `\assert{x:\readable~T,y:\writable~T,z:\writable~T}`
    y = z;
    /* actual concurrent work with x, but not y or z */
    `\assert{x:\readable~T,y:\writable~T,z:\writable~T}`
\end{lstlisting}
The single local variable assignment is enough to prevent parallelization (as the full body of a thread) via \textsc{T-Par} even though it will never cause a data race in the heap, because \lstinline|y| and \lstinline|z| are typed as \writable but \textsc{T-Par} forbids \writable references in a thread's initial type environment.
Every sound type system will reject some semantically valid code, but this example seems particularly innocuous.

\begin{figure}[t]
  \begin{center}
  $
  \begin{minipage}{0.14\textwidth}
  \begin{tikzpicture}
  \node(Top){$\mathsf{HeapWrite}$};
  \node(A)[below of=Top]{$\mathsf{NoHeapWrite}$};
  \draw(A)--(Top);
  \end{tikzpicture}
  \end{minipage}
  \qquad
  \inferrule[T-VarAssign]{x\in\Gamma \\ y \in\Gamma }{\Gamma\vdash x = y : \mathsf{unit} \mid \mathsf{NoHeapWrite}}
  \qquad
  \inferrule[T-FieldAssign]{\Gamma(x)=\mathsf{writable}\;C \\ \tau~f\in\mathsf{Fields}(C) \\\\ \Gamma\vdash e : \tau \mid \chi}{\Gamma\vdash x.f := e : \mathsf{unit} \mid \mathsf{HeapWrite}}
  $
  \end{center}
  \caption{An overly-simple effect system (excerpt) that could parallelize a local assignment of writable variables.}
  \label{fig:simple}
  \end{figure}

Consider for contrast the overly-simple effect system and rules in Figure \ref{fig:simple}.  There are two effects, \textsf{NoHeapWrite} $\sqsubseteq$ \textsf{HeapWrite}.  Every primitive expression that is not a heap write is given effect \textsf{NoHeapWrite} (notably, variable assignments), the expression that performs a write into the heap has effect \textsf{HeapWrite}, and compound expressions' effects are simply the least upper bound of the subexpressions' effects.  This way, any expression containing \emph{any} heap write would be given effect $\textsf{HeapWrite}$.  The line of code above would have effect \textsf{NoHeapWrite} --- which implies it could be parallelized without a data race.

Clearly this toy example will not scale up to real imperative programs (it likely won't handle the ``actual concurrent work'' assumed in the example), but it is still instructive because it already highlights the use-mention distinction:
the code above \emph{mentions} the writable references, but does not \emph{use} them in a way relevant to the property of interest (heap mutation).

Thus the fact that capability-bound-based reasoning does not inspect the internals of an expression is a strength in that it reduces complexity, but also a weakness because it inherently loses precision.

It is worth briefly noting that there exist reference capability systems where some references are usable only for comparing object identity, and not for actually causing effects, as in Pony's \texttt{tag} permission~\cite{clebsch2015deny}, or much earlier in Boyland et al.'s unifying framework for reference capabilities~\cite{boyland2001capabilities}. Such restricted references remain useful for code without permissions to invoke operations implemented in code with permissions~\cite{potanin2013your}.  In Pony, such references are permitted to enter recover blocks, because they do not affect the capability-bounded reasoning: they are references that do not act as capabilities (for the mutation effects addressed by Pony).

Other kinds of related, but different, distinctions have appeared in the literature on object capabilities. Miller~\cite{RobustComposition} and later Drossopoulou et al.~\cite{drossopoulou2016permission} distinguish permission as direct access to an object (to invoke its methods), and authority as the ability to cause effects on an object. Drossopoulou et al.~\cite{drossopoulou2016permission} showed that in general such notions of permission do not imply authority (a direct reference to an object with only pure methods grants permission, but not authority, over that object), and authority does not imply permission (invoking a method may cause mutations to an object the caller lacks direct access to). This distinction is further related to distinguishing permission (or authority) in a given program state from the permission (or authority) obtainable via further execution, either of a specific program, of any program adhering to some behavioral specification, or of any possible program.  The use-mention distinction somewhat resembles the distinction between eventual permission (for a given program) and behavioral permission (roughly, for all programs preserving the typing discipline, which due to types controlling permissions is also similar to Miller's notion of a topology-of-permissions based bound on authority), which also touches upon the distinction between what a program might actually do based on its code versus what it may have (or obtain) authority to do ignoring the details of the particular program.
We propose the use-mention distinction not to supplant such analyses of capability systems, but specifically to distinguish the loss of precision capability-bound reasoning suffers in comparison to effect systems.

\subsection{Working Around Use-Mention Conflation}
That the Microsoft C\# prototype was used to write an entire operating system kernel~\cite{joeblog} and Pony is used in industry suggest that at least sometimes, this use-mention distinction is not critical.  Certainly, few developers wish to parallelize a local variable assignment alone.

There are also ways to work around this limitation when it otherwise might arise.  Reference capability type systems typically include weakening (or in the C\# case, framing) type rules, that allow variables that are \emph{not even mentioned} to be temporarily set aside and ignored, allowing capability-based reasoning to be applied more locally.  
\begin{mathpar}
  \inferrule*[left=T-Weakening]{
    \Gamma\vdash e : \tau
  }{
    \Delta,\Gamma\vdash e : \tau
  }
  \and
  \inferrule*[left=T-Frame]{
    \Gamma\vdash C \dashv \Gamma'
  }{
    \Delta,\Gamma\vdash C \dashv \Delta,\Gamma'
  }
\end{mathpar}
Both rules simply state that if an expression or command is well-typed with certain variables, then it remains well-typed (with the same type) in the presence of additional variables.
Often this is enough to side-step conflation of use and mention: operations like the problematic local variable write above can frequently be refactored to a separate part of the program (e.g., before or after introducing concurrency), and this is arguably better coding style anyways.

Consider a variation on the recovery example:
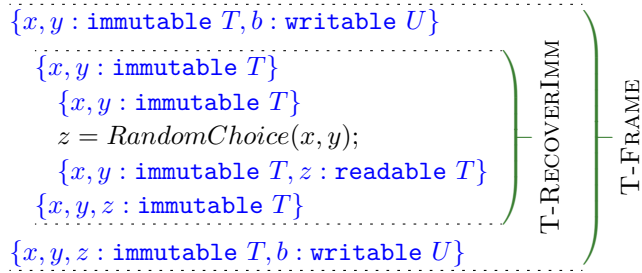
\[
\proofoutstep{x,y:\immutable~T,b:\writable~U}{
  \proofoutstep{x,y:\immutable~T}{
      \specline{x,y:\immutable~T}\\
      z = RandomChoice(x,y);\\
      \specline{x,y:\immutable~T,z:\readable~T}
  }{\textsc{T-RecoverImm}}
  {x,y,z:\immutable~T}
}{\textsc{T-Frame}}
{x,y,z:\immutable~T,b:\writable~U}
\]

This is the same code as the previous recovery example, but type-checked with an additional variable \lstinline|b| in scope with a \writable permission.  The initial type environment would fail the \textsf{IsolatedOrImmutable} check in \textsc{T-RecoverImm} because \lstinline|b| is \writable.  However framing away the extra \writable variable that is not needed in the recovery region (i.e., instantiating \textsc{T-Frame} with $\Delta=b:\writable\;U$) allows recovery to be used with an environment containing only \lstinline|x| and \lstinline|y|, both \immutable.
Thus while context-bounding risks losing precision due to the inability to distinguish use and mention, this weakness is tempered in a subtle way by the most humble of type system rules.  An under-appreciated aspect of these rules in type theories is that they imply the ``extra'' variables in $\Delta$ are \emph{definitely not used} by the expression at hand.\footnote{This is slightly surprising in contrast to separation logic, where the equivalent framing rule is (rightly) viewed as a powerful reasoning principle~\cite{reynolds2002separation}.}

It is known that removing structural rules like weakening leads to very different type theories (substructural type theories~\cite{walker2005substructural}), but we believe we are the first to remark upon this interplay between \emph{weakening} and the precision of context-bounded reasoning as a general phenomenon, rather than simply exploiting it.
Unique, linear, and affine capabilities all typically rely on restricting a different structural rule (\emph{contraction}) that permits multiple uses of the same variable (including in the aforementioned read-only reference systems).

Another more unique use of structural constraints and capabilities is the work of Giannini et al.~\cite{giannini2019flexible}, who extend the expressivity of the C\# dialect and Pony's recovery. Those languages require strict lexical nesting of recovery blocks, which can make some sophisticated uses of recovery difficult to write. Giannini et al.\ modify the structure of contexts to track multiple sets of variables for recovery simultaneously (keeping them separated), allowing a typing derivation to switch between active sets for different expressions, without any particular nesting order.  They motivate this extension from a very pragmatic point of view, but their enhancement is essentially enriching contexts with additional structure typical of a substructural logic or type system, with their new rules playing the role of novel structural rules that permute the context to swap active and ``inactive'' portions.  They noticed an interplay between structural rules and reference capabilities in a particular context, but did not highlight it as a general issue.  Still, the general issue and their result suggest deeper investigation of the interactions between capabilities and structural rules is warranted.

\subsection{The Limits of Workarounds}
Ultimately, even with the subtle benefits of weakening, the question of whether the use-mention distinction is important depends on the specific problem at hand.
For safe parallelism and method purity, the past few years have strongly suggested that the use-mention distinction is not a serious problem. Since capability-based reasoning about those effects is usually powerful enough, it is usually preferable to a full effect system due to its comparative simplicity (we see the alternative in Section \ref{sec:effects}).

Contrast this against another problem: preventing any thread other than the distinguished UI event loop thread from directly updating objects representing the UI --- considered an error in most UI frameworks, often resulting in program termination if a program violates this discipline.  
In prior work, we
proposed an effect system~\cite{ecoop13} that prevented such errors. Like the reference capability examples mentioned earlier, this has also seen adoption in industry (through Stein et al.'s clever extensions~\cite{Stein:2018:SSP:3238147.3238174}), offering some evidence that this was a good design decision. 

A key part of the work was distinguishing which objects had UI-related methods and which objects did not.  This was delineated in the type system using a type qualifier --- the same type of machinery used to manage reference capabilities --- but the actual analysis relied on an effect system.  
Because the qualifiers could be interpreted as capabilities (a thread cannot access UI elements if it holds no references to UI objects), a plausible alternative to an effect system would have been to use a context restriction on code that ran on background threads (those that should not update the UI directly): forbid them access to UI-related objects, by a rule similar to the safe parallelism rule shown earlier.  
This work was carried out shortly after work on the C\# dialect, in parallel with a related reference capability system~\cite{pldi13} refining the notion of read-only references. As a result, we considered this approach during the design of what became an effect system.

But the challenge is this: 
the details of how background threads notify the UI of completed work.  
Consider this typical sequence of steps in a user interface.  When the user clicks a button, an event handler is triggered on the UI event loop thread to handle the input.  If the work to be done is expensive, then rather than blocking the UI thread, the handler offloads work to a background thread.  Running work on the background thread will allow the UI to respond to other inputs while the work is ongoing. But once the work is done the display must be updated with the results. Background threads are forbidden from directly updating the UI themselves, for a variety of reasons discussed elsewhere~\cite{ecoop13}.
So when the work is completed, the code executing on the \emph{background thread} must somehow trigger an update to occur on the UI thread to indicate completion and/or display the results.

In all current UI frameworks, this occurs by permitting the background thread to hold (mention) a reference to UI elements, and send them in a closure to the UI thread --- which then executes the code, using the reference to update the UI.
Figure \ref{fig:uicode} gives a concrete example of this.
\begin{figure}
\begin{lstlisting}[numbers=left]
  final @UI JLabel label = ...;`\label{ln:jlabel}`
  new Thread() { // $\leftarrow$ Captures label reference
    public void run() {`\label{ln:run}` // $\leftarrow$ label reference in scope
      // do really slow computation
      Display.asyncExec(new @UI Runnable() { // $\leftarrow$ Captures label reference again
        public void run() {
          label.setText("Complete!");`\label{ln:invoke}` // $\leftarrow$ Use on UI thread
        }
      });
    }
  }.run();
\end{lstlisting}
\caption{UI event handler code spawning a background thread that sends code back to the UI thread.}
\label{fig:uicode}
\end{figure}
The \lstinline|JLabel| on line \ref{ln:jlabel} in Figure \ref{fig:uicode} is a UI element that should only be used on the UI thread.  But the background thread code (the \lstinline|Thread.run| implementation starting on line \ref{ln:run}) holds a reference to the label through the expensive work, which is then passed back to the UI thread inside a \lstinline|Runnable|, whose body (line \ref{ln:invoke}) is then safely invoked from the UI thread.
Preventing the flow of any \lstinline|@UI| object references into background threads would reject this code --- and essentially all code written for existing UI libraries.
In this case, an effect system was required to distinguish use and mention.

The use-mention distinction also arises in a second form for this problem: existing code mixes methods that should run on background threads in the same classes as methods than must run on the UI thread.  
Arguably this could be recast as a granularity issue --- splitting capabilities into those granting UI method rights and those not granting UI method rights, following the compatible aliasing approach we discuss later, could work.  But in that case it leads to capability types that are more complex than the effects --- the capabilities would need to track sets of permitted methods, while there are only two effects in 
the
solution (plus effect variables for effect polymorphism): \lstinline|@SafeEffect| $\sqsubseteq$ \lstinline|@UIEffect|.

\subsubsection{Counterarguments}
One possible objection to the above is 
that the problem above may be avoidable through use of different abstraction principles, such as defining the \lstinline|Runnable| above in a context with the \lstinline|JLabel| in scope, applying some variant of an anti-frame rule~\cite{pottier2008hiding} --- a formalization of information hiding, in this case encapsulating a capability inside the \lstinline|Runnable| --- to encapsulate the reference, and then defining the thread separately such that it cannot even (directly) mention the \lstinline|JLabel|.  However, this alone simply inverts the problem with use-mention distinctions: rather than treating mention as use, it hides both!  To ensure the background thread does not call the \lstinline|run()| method that accesses the label, it is necessary to prevent use (calling). To allow the functionality it is necessary to still allow the thread to pass the \lstinline|Runnable| to \lstinline|Display.asyncExec|.  To permit one without the other requires another distinction of use and mention --- which we would argue, is an effect system. In addition, such an approach would also prohibit background thread code from, for example, preparing a list of objects to update on the UI thread, which inherently requires the ability to mention the UI object references for storage.

A potentially stronger counterargument might stem from claiming that the difficulty with context bounding above stems from conflating capabilities with references, as all reference capability systems do.  This conflation means that capabilities can be stored in the heap.  In contrast, static capabilities divorced from data may permit additional separation: the UI thread might possess a static capability that it keeps, and UI-sensitive operations (methods) should require (and return) this unique capability.  This does make it impossible to invoke a UI operation on a background thread!
However, we would argue that this is essentially an effect system: \lstinline|@UIEffect| can be read as marking methods that require and return the hypothetical separate capability.  We are not alone in this view.

Walker et al.~\cite{Walker:2000:TMM:363911.363923} give a translation from the region calculus of Tofte and Talpin~\cite{Tofte1994Regions,tofte1997region} to a calculus of static capabilities (independent from values), and note that for this class of capabilities the distinction is in some ways a subjective difference between analyzing the behavior of code (as an effect system or monadic approach might) or dictating up front what the permissible actions are (the capability view).

More recent work on capability-based effect systems similarly takes the explicit view that capabilities grant permission to cause effects, leading to systems that restrict effects by restricting the flow of capabilities.  Liu et al.~\cite{liu2020stoic,liu2016} propose distinguishing \emph{stoic} functions as those that do not capture capabilities (directly or indirectly), and obtain stoic functions purely by capability-bounded reasoning: all functions are initially typed as possibly capturing, and a function that is well-typed in a context with no capabilities (or capability-capturing closures) can be downcast to a stoic function type (akin to recovery), which means any effects of the function then appear explicitly in its signature as capability arguments, akin to a latent effect (taking the capability as an argument does not oblige the function to use it directly). Careful use of stoic functions could be used to ensure background thread code does not capture the hypothetical UI capability, making the distinction between the two effects of interest equivalent to whether or not code accepts the UI capability as an argument.  Liu et al.\ refer to program changes to pass capabilities instead of capturing them as ``making their effects explicit.''
Osvald et al.~\cite{osvald2016gentrification} explicitly equate the capabilities required for a method with method effects, following Marino and Millstein's generic effect framework~\cite{marino09} that explicitly formulates effects as sets of capabilities.

\section{Effects, Naming, and Invariants}
\label{sec:effects}
Given the fact that effect systems can handle the use-mention distinction, why would we ever use only capabilities to bound behaviors in a static system?
The main \emph{technical} reason to choose capabilities is that they permit reasoning about effects for code that is not inspected, as in precompiled library code when retrofitting a type system, or dynamically loaded code.
But in the case that all code is compiled with a tool performing the same analysis (supporting separate compilation), this advantage is less important.  Why would we choose capabilities over effects in this case?

The answer is informal and subjective: simplicity.  Simplicity when capabilities are adequate in practice is a compelling answer for many reasonable people. But the previous section gave an example where an effect system not only handled the use-mention distinction, but was also simpler than a plausible capability-based approach.
It turns out, simplicity often favors the other direction.
Effect systems excel at reasoning about the behavior of individual sections of code --- but not at reasoning about the behavior of all code at the same time on specific shared objects with many different names.  In short, effect systems struggle to retain simplicity while enforcing invariants, particularly when they must relate multiple names to multiple entities (which is necessary to ensure multiple uses are similar).

\subsection{A Thought Experiment: Replacing Reference Immutability with Effects}
Consider, as we did,
designing an effect system that accepts precisely the same programs as a reference immutability system. For simplicity let us consider ReIm~\cite{Huang2012}, which has only mutable and transitively read-only references --- no uniqueness, and no absolute immutability.  The type rules for this system are fairly straightforward: they extend the standard class-based object-oriented type system rules to include the qualifiers in the subtyping relation, and beyond this administrative ``plumbing'' the main changes are the same one common to all deep reference immutability type systems:
\begin{itemize}
  \item The rule for type checking field writes requires the reference to the modified object to be \writable.
  \item The rule for field reads ensures that if the base object reference used for a field read is \readable, then so is the result, regardless of the permission in the field declaration.
\end{itemize}
As a consequence of these rules, for a program to follow a path through the heap to perform a write, every reference traversed along that path (local variable and field type alike) must be \writable.

An effect system with the same precision in terms of \emph{which references are used (transitively) for mutation} is quite complex.
Assuming all local variables are let-bound (i.e., final, and cannot be rebound) for simplicity, indicating that a variable was used directly for writing is straightforward: 
\[
  \inferrule{
    \Gamma(x)=T\\
    \tau\in\mathsf{Fields}(T)\\
    \Gamma\vdash e : \tau \mid \chi
  }{
    \Gamma\vdash x.f := e : U \mid \{\mathit{wr}(x)\}\cup\chi
  }
\]
This rule simply takes the type $\tau$ and effect $\chi$ of the right hand side, and adds to it an effect indicating the base reference $x$ was used for writing.
The challenge arises when reconciling external and internal variables.  Consider:
\[\mathsf{let}~x=e_1~\mathsf{in}~e_2\]
If $e_2$ contains a write through $x$, then $e_2$'s effect should include $\mathit{wr}(x)$, indicating that $x$ is used as if it were mutable.  But outside the body of this \textsf{let}, $x$ is meaningless\footnote{Or worse, means something else if it was shadowing another $x$.} --- what it refers to depends on $e_1$, and in general may refer to one of several objects (e.g., if $e_1$ involves a conditional or heap dereference).  A sound effect system would need to take any effects on $x$ and conservatively assume they could occur for any of the objects $e_1$ may evaluate to.  But this then requires the effect system to reason about may-alias relationships --- possible, but tricky, since this in turn requires naming sets of objects in the heap in a precise manner.  Essentially, an effect system approach collects aliasing and use information and propagates it outwards to be reasoned about wholesale.  For a transitive reference immutability system like ReIm, this information would also need to track origin information: it is possible that $x$ itself may never be used for writing in $e_2$, but some other reference, obtained by reading through $x$ could be --- and in that case, $x$ would need to be indicated as usable for (transitive) write access as well.

One could consider extending this experiment to more nuanced systems of read-only references.
We considered such an experiment ourselves after working on the UI threading effect system, trying to build a precise effect system analogue of the C\# reference immutability system;
the naming and usage information for an effect system approach to that language seems to grow even faster than for ReIm.
The same extrapolation applies to related systems like Pony~\cite{clebsch2015deny} and L42~\cite{giannini2019flexible}.

In this case using an effect system seems highly undesirable, and prone to significant complexity.
What changed from the UI threading effect system?  In this thought experiment, we considered a system where access paths through the heap are important, and object identity is important.  For the UI threading case, neither of those are true.
A diligent student of the literature on effect systems might point out the similarities between the considerations for let-binding above and the \lstinline|letregion| construct in calculi for region-based memory management~\cite{talpin1992polymorphic,tofte1997region}. These calculi have effect systems with similar read and write effects on a per-\emph{region} basis, rather than per object, and the effects are read and write behaviors to specific regions.  This separation from naming individual objects or tracking access paths is a substantial simplification.  The case of a region name being limited to a specific lexical scope also arises for \lstinline|letregion|, but there the region that is undefined outside that scope simply doesn't exist --- nor do any data or types that might depend on it --- because the binding construct is also the (de)allocation construct, and typing rules for \lstinline|letregion| forbid the appearance of the bound (then deallocated) region in the construct's result type. 
Object- and reference capability systems tend to be used for situations involving one or both of these features that lead to more complex effects --- object identity and heap paths.

\subsection{Global Invariants via Local Capabilities}
Capabilities, on the other hand, allow this kind of reasoning to be handled purely locally, usually without naming issues or explicit tracking of access paths.
Type contexts, along with the field type look-ups typical in type systems for OO languages excel at identifying sets of objects used similarly, because they actually \emph{force} sets of objects to be used similarly --- the type system will statically ensure that all values dynamically bound to a certain variable (or field) are used at the same type.  When absolute similarity is problematic, polymorphism over types or permissions is possible~\cite{dietl07,oopsla12,lietar2017formalizing}.
This is important because these points of the system --- variable and field types --- already conflate types of different objects in standard type systems.
So tying capabilities to variable and field types essentially enforces a kind of invariant: it conflates capabilities in the same places a traditional type system already conflates basic types.  As a result, this leads to little additional friction for developers already using a typed language.
Effect systems such as the hypothetical effect version of reference immutability must somehow reconstruct this sort of conflation that comes for free when the effects are restricted \emph{by} the type context.

Static reference capability systems of recent years also all carry a notion of \emph{compatibility} between references/capabilities. In many static reference capability systems, each reference permission comes with not only restrictions on how it is used, but restrictions on how \emph{aliases} are used.
These systems maintain a global invariant that for any two aliases, the permissions granted via one reference are a subset of the interference assumed by the other, in both directions.
The early papers on rely-guarantee references~\cite{pldi13}, rely-guarantee protocols~\cite{militao14}, and Pony~\cite{clebsch2015deny} give particularly thorough accounts of this.
This notion of compatibility between aliases is imposed any time references are duplicated, and in the case of systems like Kappa~\cite{CastegrenW16}, joined as well.  

Preserving compatibility between aliases can also be done locally, without name binding issues.
In each case, one type $A$ may be split into two others $B$ and $C$ if:
\begin{itemize}
\item $B$ and $C$'s combined capabilities do not exceed $A$'s original capabilities for modification, and
\item $B$ (resp. $C$) assumes at least as much interference as $A$ assumed
\item $B$ (resp. $C$) assumes at least as much interference as $C$'s (resp. $B$'s) capabilities provide.
\end{itemize}
As a concrete example, consider the rely and guarantee components of a rely-guarantee reference~\cite{pldi13,toplas17}, which specify binary relations constraining what modifications that reference may be used for (the guarantee) and what its aliases may be used for (the rely).
A reference of type $\mathsf{ref}\{\mathbb{N}|>5\}[\le,=]$ refers to a natural number strictly greater than 5, assumes aliases may increment the number (any time an alias modifies the stored value, the old value must be $\le$ the new value, and typing may \emph{rely} on this fact), and may only be used for reading (or non-modifying writes; new values must be $=$ the old value, and the type system must \emph{guarantee} uses obey this restriction).
This may be split into two copies of itself (it is \emph{reflexively splittable}), because none of the three (original and the two split copies) permits writes, but all would tolerate increments through aliases.  
Moreover, because the predicate on the referent (that it is greater than 5) is preserved by the guarantee (equality), this check on reference splitting ensures the predicate will be preserved by \emph{all} possible references, with only point-wise checks every time a new alias is created.
In contrast, a reference of type $\mathsf{ref}\{\mathbb{N}|>5\}[=,\le]$ may be used for incrementing, but assumes all aliases are read-only.  So it may not be duplicated na\"ively: each copy would assume it was the \emph{only} reference that could be used for increments.
This permits some very granular reasoning about side effects, without a full effect system (though again, not distinguishing mention and use). 

As one could imagine, extending our thought experiment of a purely effect system replacement for ReIm to a system like this would produce very complex effects, adding constraints from these binary relations into effects dealing with naming and aliasing. By enforcing this restriction on duplicating references, the type system can ensure the value stored in that reference remains greater than 5 without explicitly tracking where the aliases go or when they are used.

In the ``reference immutability'' family of read-only reference type systems~\cite{oopsla12,clebsch2015deny,huang2012inference,Huang2012,zibin07,zibin10,tschantz05}, compatibility typically requires no special care --- the shape of the permission subtyping relationships already ensures any duplication preserves compatibility (setting aside unique references). \readable and \writable references assume aliases may mutate the referent, and while \immutable references assume no aliases may mutate it, they also do not grant permission for mutation, so duplication is not problematic.

In other systems, the changes remain relatively local following the general argument above. Rely-guarantee references~\cite{pldi13,toplas17} use a notion of type splitting, $\Gamma\vdash \tau \prec \tau' \divideontimes \tau''$ to check that when a value (particularly one containing references) is duplicated, it can be split into compatible types $\tau'$ and $\tau''$.  
It generally recursively checks splitting, bottoming out at the reference splitting rule, which looks somewhat complex but merely formalizes the three aspects of compatibility above (plus preservation of predicates):
\newcommand{\reft}[4]{\mathsf{ref}\{#1\mid#2\}[#3,#4]}
\begin{mathpar}
  \inferrule*[left=\scriptsize Ref-$\divideontimes$]{
    \Gamma\vdash\reft{b}{\phi'}{R'}{G'}\\
    \Gamma\vdash\reft{b}{\phi''}{R''}{G''}\\
    \emptyset\subset \db{G'}\subseteq \db{R''}\\
    \emptyset\subset \db{G''}\subseteq \db{R'}\\
    \db{G'}\cup\db{G''}\subseteq\db{G}\\
    \db{R}\subseteq\db{R'}\\
    \db{R}\subseteq\db{R''}\\
}{
    \Gamma\vdash\reft{b}{\phi}{R}{G}\prec\reft{b}{\phi'}{R'}{G'}\divideontimes\reft{b}{\phi''}{R''}{G''}
}
\end{mathpar}
This formalizes splitting type $A$ into types $B$ and $C$ ($\Gamma\vdash A\prec B\divideontimes C$) when all are rely-guarantee references.  Beyond checking that the new types $B$ and $C$ are well-formed, it checks that $B$ and $C$'s combined capabilities (guarantees) do not exceed $A$'s ($G'\cup G''\subseteq G$), that $B$ assumes at least as much interference as $A$ ($R\subseteq R'$), and that $B$ tolerates interference from $C$ ($G''\subseteq R'$) (plus the symmetric checks on $C$).

This splitting check is inserted into a couple obvious locations in static reference capability systems, wherever new aliases may be created --- variable reads, memory reads, and parameter passing.  Rely-guarantee protocols~\cite{militao14,Militao16Composing} do a form of model checking to check compatibility in the same places.
Kappa~\cite{CastegrenW16} has a similar notion of packing and unpacking composite capabilities.
Maintaining this compatibility invariant with only local checks means that the concurrent versions of these systems~\cite{Militao16Composing,CastegrenW16,toplas17} no longer require explicit bounding checks for concurrency --- simply splitting well-formed type contexts (and certain assumptions about the granularity of interleaving) is sufficient for safety.
And because the combined permissions of the two new references cannot grant more authority than the original's permissions, any invariant enforced by the original is enforced by both new references as well.\footnote{In systems that permit recombining reference capabilities~\cite{CastegrenW16,CastegrenW17,Militao16Composing,militao14}, the new reference may grant more permissions that the two original pieces, but the system maintains that rejoining previously-split references never grants more authority than the original.}
Typestate managed via permission~\cite{naden2012type,Garcia2014FTP} has analogous checks.

This discussion, however, is abstracted from concrete use cases.  And it is worth asking whether some particular aspects of reference immutability, particularly the transitive variants, might make the problem worse than it could be (though we didn't get that far above).

\subsection{Invariants for JavaScript, Instead of Effects}
We previously encountered the challenges involved in maintaining global invariants with effects when designing a
type system to enable efficient ahead-of-time compilation of JavaScript~\cite{oopsla16}.
The goal was to allow JavaScript to be run on embedded devices, faster than via an interpreter, but with lower memory footprint than a JavaScript JIT (which in addition to keeping the compiler in memory, keeps multiple versions of the code).
The core idea behind the type system was to use types to rule out JavaScript behaviors that are especially difficult to optimize at compile time --- those that would seem to require a JIT to execute efficiently --- while permitting some of JavaScript's (in)famous flexibility that did not seriously interfere with compilation.  JavaScript's semantics are full of cases that are difficult to compile efficiently ahead of time, but we will focus on one particularly tricky case that pushed the team towards capabilities.

One aspect of JavaScript that makes it particularly difficult to optimize is the fact that object layouts are not fixed --- fields may be added or removed dynamically.
This means the typical approach to compiling field accesses in a language like Java or C --- emitting a constant-time access to a statically-known offset from the object's base pointer --- does not work in general.  
Fortunately, a significant amount of JavaScript code is reasonably well-behaved and does not add fields once an object is fully initialized. But because normal JavaScript will silently create fields if a program writes to one that doesn't exist, it is easy to do this unintentionally.

Consider the code in Figure \ref{fig:jscode}.
\lstinline|F| is a (pre-ES6) constructor.  Calling \lstinline|new F()| allocates a new object, sets \lstinline|F.prototype| as that object's \emph{prototype} (source of inherited properties), and executes the code of the function \lstinline|F| with that new object as the receiver. In JavaScript, if a field is read on an object, but does not exist there, the runtime checks for that field in the object's prototype.  If it is there, it returns the value from the prototype.  Otherwise the runtime checks the prototype's prototype, and so on, until the field is found or there are no more prototypes.  A field write, however, always writes to the immediate referent, and never consults the prototype chain.
This makes subtle mistakes possible.  
The call to \lstinline|f.inc()| increments the field \lstinline|x| in \lstinline|f| as expected; \lstinline|inc| is found in the prototype object, invoked with \lstinline|f| as the receiver, and the write in that method writes to \lstinline|f|.
The last line of Figure \ref{fig:jscode} invokes the method on the \emph{prototype}, however, which is probably not supposed to have an \lstinline|x| field at all.  In standard JavaScript runtimes, this would run without error:
reads of undefined fields return a special \lstinline|undefined| value, which is coerced to a number (really, \lstinline|NaN|) by addition, and the increment then writes to f, which will result in the runtime dynamically adding the field.
But \lstinline|F.prototype| is intended to be the equivalent of an abstract class --- all methods, no data.
For the purposes of ahead-of-time compilation, this would be a problem to avoid.

\begin{figure}
\begin{lstlisting}[morekeywords={function,var}]
function F() {
  this.x = 0
}
F.prototype.inc = function() { this.x++; }
F.prototype.count = function() { return this.x; }
F.prototype.incAndCount = function() {
  this.inc();
  return this.count();
}
/* construct a new F instance, and increment its x field */
var f = new F(); // f.x == 0
f.inc();         // f.x == 1
/* add the field x to F.prototype */
F.prototype.inc();
\end{lstlisting}
\caption{Violating fixed-object layout.}
\label{fig:jscode}
\end{figure}

The heart of the problem above is that the \lstinline|inc| method writes to \lstinline|this.x|, and therefore should only be executable on objects that (should) have a field \lstinline|x| before the call.  The last line of code should then be rejected 
because it calls \lstinline|inc| on an object missing required fields.
The actual system design included many other issues, but this problem could be viewed as the defining challenge for the system: if all objects were guaranteed to have fixed object layout, then a runtime system incapable of dynamic field addition and removal could still preserve the original program semantics.

Building a type system for a dynamic language essentially always requires
structural types (i.e., record types with width subtyping~\cite{cardelli1991operations}), which enumerate which fields were present in each object, leading to types like
\[\{x:\mathsf{number}, y:\mathsf{number}, m:()\rightarrow\mathsf{number}\}\]
indicating two numeric fields and a method returning a number.\footnote{In this exposition, we will only consider methods, even though the full system supported functions as well.}
Initial work on the project~\cite{choi2015} also made clear a need to
distinguish definitely-local fields (like \lstinline|f.x|) that could be written safely, and possibly-inherited fields (like \lstinline|f.inc|) --- field accesses to the former can be compiled more efficiently than the latter.
This leads to split object types of the form $\{r\mid w\}$, where $r$ contains the types of \emph{readable} fields known to be present somewhere (locally or inherited), and $w$ contains the types of \emph{writable} fields known to be present on the immediate referent.

We can explore another thought experiment, which is actually a reproduction of the original trajectory in designing this as an effect system, prior to correcting to a capability system.
Initially, it appeared\footnote{What follows reflects a personal view of what appeared ``obvious'' at different points in time, and the actual design process the present author engaged in; we do not mean to suggest our coauthors were predisposed to the same mistakes.} we should view the problem in terms of which fields of each object were accessed (in which ways) by each section of code.
In hindsight, we can concisely state that the goal was to ensure each object had a fixed object layout, and that all references to each object collaboratively maintained that fixed layout as an invariant, as alluded to in the previous subsection.  
Both of these are correct points of view, but they lead to very different system designs.

\subsubsection{The Effect System Approach}
An early approach to handling the problem in Figure \ref{fig:jscode} was
an effect system tracking which fields of the receiver were written by each method.  In this case, the problematic call above is rejected by the draft rule \textsc{T-MCallSketch}: \lstinline|F.prototype|'s type does not include \lstinline|x|, while \lstinline|inc|'s effect would indicate it would write \lstinline|this.x|.
\begin{mathpar}
  \inferrule[T-MCallSketch]{
    \Gamma\vdash e : \{ r \mid w \} \mid \chi\\
    m:(\tau_1,\ldots,\tau_n)\xrightarrow{\chi_m}\tau\in(r\cup w) \mid \chi_i\\
    \forall i\in 1..n\ldotp \Gamma\vdash e_i : \tau_i \mid \chi_i\\
    \chi_m\subseteq w
  }{
    \Gamma\vdash e.m(e_1\ldots e_n) : \tau \mid \chi\cup\bigcup_{i\in 1..n}(\chi_i)
  }
\end{mathpar}
In particular, the final antecedent (the subset check) would fail.

Going even slightly beyond this example, however, quickly pushes this idea into unwieldy territory, because
this requires tracking not only
presence or absence of object modification as in the previous thought experiment, but also which parts of an object were modified.
Objects also sometimes pass the receiver as an argument to methods of \emph{other} objects (notice that if one of the parameters passed in \textsc{T-MCallSketch} is the receiver, this --- unsoundly --- does not affect the overall effect). So to track the correct set of receiver field writes for a method containing \lstinline|foo.bar(this)|, it becomes necessary to track which fields \lstinline|foo.bar| actually writes to on its (initial) first argument --- accounting for subsequent aliasing and transitive calls within \lstinline|foo.bar| as well.

But the trouble does not end there, as it did in the ReIm effect system thought experiment above.
Reference immutability type systems (and reference capability systems in general) only articulate constraints on interface components --- the receiver, method parameters, and return value --- and need not explicitly describe internal behaviors, keeping the types relatively simple.
These effects, however, expose internal implementation details of objects, like ``private'' field names.  For examples like Figure \ref{fig:jscode} alone, this abstraction violation is merely uncomfortable.  But it quickly becomes a technical problem. 

Notice that instances of Figure \ref{fig:jscode}'s \lstinline|F| implement a structural interface with methods to increment a counter and get its current value.  
Assuming the split object types outlined above,
\lstinline|f| can be given a concise type:
\[\{ inc:()\xrightarrow{x}() \mid x:\mathsf{number}\}\]
This type says the object has (possibly-inherited) fields \lstinline|inc| and \lstinline|get|, and a local field \lstinline|x|.
If another object \lstinline|g| implements the same interface, but uses internal field name \lstinline|y| to store its count, it would have type:
\[\{ inc:()\xrightarrow{y}() \mid y:\mathsf{number}\}\]
Now we have a problem: what are the effects of these methods in the least common supertype of these types, which we would need to store \lstinline|f| and \lstinline|g| in the same local variable or pass them to the same methods?
The increment method's effect mentions \lstinline|x| in \lstinline|f|'s type, while the effect of \lstinline|g|'s increment method mentions \lstinline|y|.  The effects are incompatible.

Depth subtyping on mutable records is unsound in general, but the methods are in the read-only part of the object (since they are inherited), so depth subtyping is sound for them.
This means that for the \lstinline|inc| method, using subtyping to over-approximate the actual effect of each method is sound, so the least upper bound of the incrementing interfaces could then be:
\[\{inc:()\xrightarrow{x,y}() \mid \}\]
This combines width subtyping (which drops fields that do not exist in both objects) with depth subtyping on the read-only fields.
This is a meaningful upper bound: the latent effect over-approximates both implementations' effects.  But \lstinline|x| and \lstinline|y| do not appear in this type, so 
checking that such an object contains all fields mentioned in the method effects in order to type-check a method invocation would fail --- and in fact, neither object has \emph{both} field \lstinline|x| and field \lstinline|y|.

We can resolve this, perhaps, by existentially quantifying over the particular field.  But since this is a general issue of representing internal state, we must also abstract over the field's type.  And of course, there's no requirement that two implementations of the same abstract interface use the same \emph{number} of fields to store their state, leading to existential quantification over \emph{rows}~\cite{wand1989type} --- essentially fragments of object types\footnote{Rows were originally used as an alternative to bounded polymorphism in object or record calculi, such as $\forall X::\mathsf{row}\ldotp x\not\in X \Rightarrow \{x:\mathsf{number},X\}\times\{x:\mathsf{number},X\}\rightarrow\{x:\mathsf{number},X\}$ as the type of a function that takes two objects with common fields including a field $x$, and returning whichever has the larger value in the field $x$.  Rows are now also used in effect systems~\cite{lindley2012row} in an analogous way, but this is orthogonal to our capabilities vs.\ effects discussion.}:
\[\exists X::\mathsf{row}\ldotp \exists W::\mathsf{row}\ldotp \{ inc : ()\xrightarrow{\mathsf{wr}(X)}() \mid W
 \}\]
This type essentially says the $\mathit{inc}$ method modifies some set $X$ of receiver fields, and existentially quantifies over locally-present fields.

But even this is not a complete solution!  Now we can again store references to \lstinline|f| and \lstinline|g| in the same storage location by making different choices for the existentials, and now no longer leak information about the names of internal fields. But we haven't solved the original problem.  We still need to know if the now-existentially-quantified row of fields written by the method is a subset of the fields actually present in the object in order to invoke the method.  This information is not only lost by width subtyping and the abstraction of the existential, but the relationship between the row and other fields the object may contain is not captured by the type.

In the more concrete case of \lstinline|f| and \lstinline|g|, their common supertype again cannot explicitly mention the presence of \lstinline|x| or \lstinline|y|, since neither field is in both objects.  This leads to further existential quantification, and bounding of row variables! To actually \emph{invoke} \lstinline|inc| through the abstract interface, we must know the written fields are a subset of the present fields.
We can embed this information by using \emph{bounded existentially quantified row variables}:
\[
  \exists W::\mathsf{row}\ldotp\exists X\subseteq W\{ inc : ()\xrightarrow{\mathsf{wr}(X)}()\mid W
 \}
  \]

But at the cost of some complexity, it seems this does offer a path to solve the original problem: each method may possibly write different subsets of local fields, 
and it seems if enough constraints are added, it should be possible to make the necessary connections to check that invoked methods only access fields that are actually present on the receiver.

Yet it is still not a complete solution.  This path can handle the increment example.  But to solve the original problem, two \emph{additional and substantial} extensions are still required.  First, there is a parallel problem with methods possibly \emph{reading} fields that may not be present in the prototype chain.  Without completing this exercise in full detail, note that because field reads and writes do not obey quite the same restrictions, handling reads effectively doubles the number of row variables and bounds (for every method signature), though the bounds for reading are slightly more relaxed than those for writing (since fields may be local \emph{or} inherited).
Reading from an inherited field is acceptable and is in fact how method dispatch commonly works in JavaScript. With the code in Figure \ref{fig:jscode}, calling \lstinline|f.incAndCount()| should be permitted, even though the body of that method, inherited from the prototype, invokes (and therefore reads) two inherited method fields.  Extending for method-read sets results in types like this one, which adds more complex constraints to deal with the fact that reading writable fields is safe:
\[
  \exists R,W::\mathsf{row}\ldotp\exists X\subseteq W\exists Y\subseteq(R\cup W)\ldotp\{ inc : ()\xrightarrow{\mathsf{wr}(X)}(),
     get : ()\xrightarrow{\mathsf{rd}(Y)}\mathsf{number},R \mid W
 \}
  \]

Second, we have not addressed the additional complication mentioned earlier: the receiver may escape a method, so tracking \emph{only} the receiver fields a method modifies is insufficient!  Consider a method body that registers the receiver for updates:
\begin{lstlisting}
Foo.prototype.reregister = function() {
  this.targetSource.registerListener(this);
}
\end{lstlisting}
If the \lstinline|registerListener| method modifies its argument (directly or by invoking methods that do so), those modifications should also be reflected in the effect of the \lstinline|reregister| method.  But the only way for this to work is if the type or effect of \lstinline|registerListener| reflects the fields it updates \emph{on its arguments}, as well as on its receiver (\lstinline|this.targetSource| in this case). This also brings in the aliasing issues discussed in the effect system reconstruction of ReIm.

As presented here, the complexity is clearly significant even before it is carried to its logical conclusion.  But at what point during this design process did it \emph{become too complex}? Can we identify a point in this design evolution where it clearly crossed the line?
The project required structural types for objects from the start, so it's hard to tell exactly which pieces of the growth above are truly necessary and which add too much complexity:
rows for instance originated in type inference for record calculi~\cite{wand1989type}, and these kinds of constraints between rows were known to be necessary to type certain kinds of programs~\cite{cardelli1991operations}.
The project goals included regular developers using the result, so inference was a requirement, which then implied rows and row constraints had a role to play.
The eventual implementation uses rows, though row \emph{constraints} are limited to type inference only and ultimately do not appear in surface types seen by developers.
Many type systems with unpleasant core complexity manage to tame some of it through convenient short-hands and careful selection of default assumptions.
So while hindsight shows this approach would have led to more complex metatheory and implementation, and probably significantly worse error reporting, the fact that this approach had some justification in its relationship to inference, and clearly exposed all of the required information, made it harder to tell when this route might have crossed the line to being unacceptably complex.

A fair question to ask at this point is also how much of this complexity stems from the particular problem at hand --- reasoning about the particular interaction of field reads and writes with JavaScript's uncommon inheritance model. Greenhouse and Boyland's work on an object-oriented analogue~\cite{greenhouse1999object} of FX~\cite{lucassen88,gifford86} (an effect system for reasoning about non-interference of program expressions) resembles early stages of the development outlined here.  They continued the FX emphasis on regions, and permitted Java classes to declare abstract regions of fields. Regions existed in a nesting hierarchy (which inspired the same structure in DPJ~\cite{bocchino09}), such as a hashtable having nested regions for keys and values to separate impacts on those parts of the structure.  Method effects were then the set of regions read or written by the method, with field names acting as special (very specific) regions. Effects could refer to specific object (e.g., the value region of a hashtable taken as a parameter), which is roughly analagous to the outline we gave for handling the \lstinline|reregister| example. As a result actually checking their effect system requires points-to information~\cite{greenhouse1999object,boyland1999mayequal}.

\subsubsection{Back to Capabilities for Invariants}
Starting from the outline above, how did we simplify the system? We can see several steps to condense the information from our hypothetical complex effect system down to the still-sophisticated, but more manageable published system~\cite{oopsla16}.
The first step was to simply impose a single upper bound on the written receiver fields, shared across all methods on that object.
Thus, object types
would
(sometimes) contain \emph{two} kinds of object types: a physical type describing the local and inherited fields (which fields are actually present, and which are writable), and a method-required type describing sufficient receiver assumptions to execute any attached methods.
This moves part of the effect information from the methods to the object type itself (and is a feature of the final system).
The published system calls the method-required portion of the type the \emph{method-accessed fields}.
Because both present and method-access fields must further split into distinctions between possibly-inherited (and therefore readable) and definitely-local (and therefore writable), this resulted at one point in four-part object types
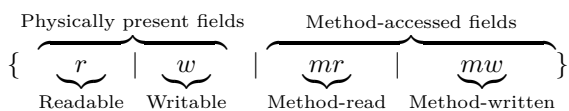
\[ 
  \{
    \overbrace{
      \underbrace{r}_\text{Readable}\mid
      \underbrace{w}_\text{Writable}
    }^\text{Physically present fields}
    \mid
    \overbrace{
      \underbrace{mr}_\text{Method-read}\mid
      \underbrace{mw}_\text{Method-written}
    }^\text{Method-accessed fields}
  \}
\]
where each variable is a row:
\begin{itemize}
  \item $r$ contains definitely-present, but possibly-inherited fields, which are safe to read.
  \item $w$ contains definitely-present, definitely-local fields, which are safe to write.
  \item $mr$ contains fields that may be read by some method, but are definitely not written by any method of the object.
  \item $mw$ contains fields that may be written by some method of the object.
\end{itemize}
The method-access fields are taken to be a single upper bound on the effect of any method on the object, dualized to describe the \emph{capabilities sufficient to execute any method on the object}.
The other fields describe the physically present fields of the object, distinguishing those that are definitely local and can therefore be written without affecting object layout.

Then using the read-write split on physical fields ($r$ and $w$) then becomes apparent as a way to summarize how a method uses its arguments --- if \lstinline|registerListener| above modifies field \lstinline|foo| of its argument, it will be reflected in the required parameter type containing a writable field \lstinline|foo| of the appropriate type, which we can interpret as a reference capability required by \lstinline|registerListener|.
Since the types in the system already needed to track which fields are on the immediate referent (and therefore, safe to write without changing field layout) and which are possibly-inherited (so safe to read, but not necessarily safe to write), this actually removes some redundancy: the physical layout information plays double-duty as both a physical description \emph{and} a capability granting read-access to present fields and write-access to local fields.  And while it again begins to sacrifice the use-mention distinction, for this problem the distinction turns out not to be critical.

``Flattening'' use information from effects into mention information in reference types (capabilities) addresses the issue of soundly tracking reads and writes.  This leaves us with two other challenges raised above: reasoning about when it is actually safe to invoke a method, and abstracting types in a way that we can invoke methods based on interfaces with different implementations.
Turning to the notion of asymmetric compatible capabilities that collaboratively enforce an invariant, we find another solution.
When deciding whether it is safe to invoke a method, it is not really relevant which particular fields are present, only that those present include the ones accessed by methods (again, informally blurring some distinctions between reads and writes).  

We can shift our view to maintaining each object as being either an \emph{abstract} object (whose methods access fields that are not present, by analogy to an abstract class), or a \emph{concrete} object with all the fields required (in the appropriate places) to safely invoke \emph{any} of its methods (since there is now only one common bound on the behavior of all methods on an object).
We can view membership in one of these sets as an invariant collaboratively maintained by all references to an object.
Given one of the ``double'' object types suggested above, the check is simple: if every field assumed writable or readable by methods (i.e., in the method-accessed fields) is actually writable or readable on the physical object (i.e., in the right partition of the physically-present fields), then it is safe to invoke methods on that object.  Moreover, once that check is performed for a given object, since the method-access field information for the object and the physical layout information should be invariant, the information about method-accessed fields can be discarded, leaving only the basic physical object type ($r$ and $w$) as important.

For example, consider \lstinline|f| and \lstinline|g| from our earlier example.  \lstinline|f| would be given (full) type:
\[ \{ inc:()\rightarrow()\mid x:\mathsf{number} \mid \emptyset\mid x:\mathsf{number}  \}\]
and \lstinline|g| would receive the analogous type mentioning $y$:
\[ \{ inc:()\rightarrow()\mid y:\mathsf{number} \mid \emptyset\mid y:\mathsf{number}  \}\]
Since the method-written set is contained in the physically local writable set for each object, \lstinline|f| can be given the simpler object type $\{inc:()\rightarrow()\mid x:\mathsf{number}\}^\textsc{NC}$, where \textsc{NC} tags the object as \emph{concrete}, indicating the check was performed when method-accessed fields were known, and aliases will ensure that check remains true. \lstinline|g| can be given the analogous type mentioning $y$, and then traditional width subtyping\footnote{Tweaked for the read/write split of fields.} lets both be given the common supertype $\{inc:()\rightarrow()|\emptyset\}^\textsc{NC}$.  This common supertype only mentions the method of interest, using standard subtyping to hide the irrelevant differences. But because it is flagged as concrete, the type system can permit the increment method to be invoked: the \textsc{NC} tag indicates the referent already satisfies sufficient invariants for any method invocation to be safe, and restrictions on how aliases are created (essentially, sound treatment of subtyping and field updates) ensure the invariant is preserved. Most people would agree this is substantially simpler than the type laden with explicit row quantification and constraints.

The only time the full double object types are required is when handling prototype objects (e.g., for initialization) or replacing existing methods.  In those cases, it is necessary to check that the method-required half of the object type is (informally) a subtype of the \emph{assumed receiver type} of a newly-installed method.  Intuitively, that method-accessed sets are an invariant of the object, and attaching a method ensures the new method preserves that invariant (i.e., does not install a method that accesses other things).  Read as capabilities, the full object types provide the extra information / permissions required to check method replacement, which takes the form of unattached methods with assumed receiver types stating the permissions required by the new method body.
Chandra et al.\ call these full types \emph{prototypal types}, and distinguish them from \emph{non-prototypal types} that carry no method-accessed fields because they can only be created from prototypal types when the check that all method-accessed fields are present succeeds.
In some cases complete objects may also be used as prototypes, so some objects may be aliased by references with dual types (prototypal) and by references with single types (non-prototypal).
The non-prototypal concrete (i.e., \textsc{NC}) types grant the capability to invoke any visible methods. The dual (prototypal) types grant the capabilities to modify prototype or method members (and carry sufficient information to actually perform the containment checks between local fields and method assumptions).

While the discussion above focused on reasoning about access to specific fields, it is worth noting that all structural object types --- including those just discussed --- form a sort of reference capability with support for static delegation (but not revocation).  If a developer wishes to pass an object to some code, but limit which methods of the object may be invoked, using width subtyping one can obtain a reference which does not mention the ``restricted'' operations, and a sound type system (and limiting reflection) ensures a callee will not

\section{Conclusion}
We have outlined what we have found to be the major trade-offs in practice between static (reference) capabilities and effect systems: choosing between simpler design and abstract reasoning principles, and handling the use-mention distinction.  We have also highlighted examples of a subtle interplay between reference capabilities and modest aspects of type systems (weakening rules and type contexts) that results in useful added expressive power in a way that has not been highlighted previously.  Lastly, we have tried to put these in context by 
explaining what breaks --- functionally, or by introducing unwieldy complexity --- when considering effect system versions of reference capability systems or vice versa, based on our personal experience facing these trade-offs while designing reference capability systems and effect systems.
We hope primarily that this will be useful to others in choosing between approaches to static reasoning, and helpful to newcomers seeking to better understand the trade-offs between these approaches.

\section*{Acknowledgments}
Many thanks are due to the audience at the OCAP 2018 workshop where these ideas were initially presented, and to the ECOOP 2020 reviewers, for helpful feedback on the ideas, presentation, and paper itself.

\bibliographystyle{plainurl}
\bibliography{csg,effects}

\end{document}